\documentclass[11pt,twoside]{article}

\usepackage{asp2010}
\usepackage{graphicx}

\usepackage{amsfonts}
\usepackage{amssymb}
\usepackage{bm}
\usepackage{dcolumn}
\usepackage{epsfig}
\usepackage{graphics}
\usepackage{graphicx}
\usepackage{hyperref}
\usepackage[latin1]{inputenc}
\usepackage{latexsym}
\usepackage{multirow}
\usepackage{rotating}
\usepackage{upgreek}

\newcommand{\mb}[1]{\mbox{\boldmath $#1$}}

\newcommand{\INC}{{\mbox{\tiny inc}}}

\newcommand{\met}{\mbox{g}}

\newcommand{\NN}{{\mbox{\tiny N}}}

\resetcounters 

\begin{document}
\resetcounters

\title{Constraining Gravity with LISA Detections of Binaries}
 \author{Priscilla Canizares$^1$, Jonathan R. Gair$^1$ and Carlos F. Sopuerta$^2$ \affil{$^1$ Institute of Astronomy, Cambridge, United Kingdom.}\affil{$^2$ Institut de Ci\`encies de l'Espai (CSIC-IEEC), Campus UAB, Torre C5 parells, 08193 Bellaterra, Spain.}}
 
\begin{abstract} 
General Relativity (GR) describes gravitation well at the energy scales which we have so far been able to achieve or detect. However, we do not know whether GR is behind the physics governing stronger gravitational field regimes, such as near neutron stars or massive black-holes (MBHs). Gravitational-wave (GW) astronomy is a promising tool to test and validate GR and/or potential alternative theories of gravity. The information that a GW waveform carries not only will allow us to map the strong gravitational field of its source, but also determine the theory of gravity ruling its dynamics. In this work, we explore the extent to which we could distinguish between GR and other theories of gravity through the detection of low-frequency GWs from extreme-mass-ratio inspirals (EMRIs) and, in particular, we focus on dynamical Chern-Simons modified gravity (DCSMG). To that end, we develop a framework that enables us, for the first time, to perform a parameter estimation analysis for EMRIs in DCSMG. Our model is described by a 15-dimensional parameter space, that includes the Chern-Simons (CS) parameter which characterises the deviation between the two theories, and our analysis is based on Fisher information matrix techniques together with a (maximum-mismatch) criterion to assess the validity of our results. In our analysis, we study a $5$-dimensional parameter space, finding that a GW detector like the Laser Interferometer Space Antenna (LISA) or eLISA (evolved LISA) should be able to discriminate between GR and DCSMG with fractional errors below $5\%$, and hence place bounds four orders of magnitude better than current Solar System bounds.
\end{abstract}

\section{Introduction} \label{introduction}

EMRIs are low-frequency GW systems detectable by GW observatories like (e)LISA. They are made up of a MBH ($M^{}_{\bullet} \sim 10^4-10^7 M^{}_{\odot}$) and a stellar compact-object (SCO) ($m^{}_{\star} \approx 1-50 M_{\odot}$). Once the SCO is captured by the MBH, the system loses energy and angular momentum and the SCO starts inspiralling driven by the emission of GWs. During this process, the geometry of the MBH spacetime where the inspiral take place is mapped in the shape of the GWs emitted. Therefore, by looking at the phase and timing of an EMRI waveform, we will be able to determine the physical parameters of the system and the structure of the MBH potential. For this reason, waveform models used for GW detection rely on the structure of the MBH spacetime and, hence, on the theory of gravity assumed. 

GR has been well tested in the weak field, but the behaviour of gravity in these weak gravitational regimes could be effectively described by GR, whereas in strong gravitational fields the true theory might deviate from GR. In that case, gravity may be described as an effective theory whose action contains higher-order curvature terms. Then, since EMRIs emit deep inside the strong MBH potential, through the detection of their GWs we can unveil the physics behind these systems and reveal the theory of gravity that governs them.

In this contribution, based on ref.~\citep{Canizares:2012is}, we focus on studying the detectability of DCSMG, a modification of GR constructed by the addition of a Chern-Simons (CS) gravitational term (also known as the Pontryagin invariant) to the action. Interest in this theory was initiated with the work of Jackiw and Pi ref.~\citep{jackiw:2003:cmo} where gravitational parity violation was investigated. In addition, DCSMG presents distinctive features that could be observed by GW detection, such as higher-order curvature terms which become important in strong gravitational fields. This work follows the conventions of ref.~\citep{Misner:1973cw}.

\section{Parameter Estimation in DCSMG} \label{formulation}

In order to study EMRIs in DCSMG, we need to model the orbit of the SCO and the GWs generated by the system in this theory, which means that we have to know the metric of the spacetime where the SCO moves. Besides from the black hole mass $M_{\bullet}$ and spin $a$, black hole metrics in DCSMG are parameterized by a single universal constant, $\xi:=\alpha^2/(\beta\kappa^{}_{\NN})$, where $\kappa^{}_{\NN}$ is the gravitational constant, and  $\alpha$ and $\beta$ are universal coupling constants that control the strength of the CS modifications. Then, $\xi$ is the only parameter responsible for the deviations between GR and DCSMG. The non-vanishing components of the DCSMG metric are ref.~\citep{Yunes2009}:
\begin{eqnarray}
{\met}^{}_{tt}  &=&  -\left(1-\frac{2M^{}_{\bullet}r}{\rho^2}\right) \label{gtt}\;,~~~~
{\met}^{}_{rr} =  \frac{\rho^2}{\Delta}\;,~~~~
{\met}^{}_{\theta \theta}  =  \rho^2 \;,~~~~
{\met}^{}_{\phi \phi}  =  \frac{\Sigma}{\rho^2}\sin^2{\theta}\\ \nonumber
{\met}^{}_{t\phi} & = &\left[\ \frac{5}{8} \frac{\xi}{M_{\bullet}^{4} } \frac{a}{M^{}_{\bullet}}
\frac{M^5_{\bullet}}{r^4}\left(1+\frac{12M_{\bullet}}{7r} + \frac{27M^2_{\bullet}}{10r^{2}}\right)- \frac{2M^{}_{\bullet}a r}{\rho^2}\ \right]\sin^2\theta\;\,,\label{DCSMG_metric}
\end{eqnarray}
with $\rho^2=r^2+a^2\cos^2\theta$, $\Delta = r^2f+a^2$, $f = 1-2M^{}_{\bullet}/r$, and $\Sigma=(r^2+a^2)^2-a^2\Delta\sin^2\theta$. 

Once we have the metric of the MBH spacetime in DCSMG, we obtain the geodesic equations ref.~\citep{Yunes2009}:
\begin{eqnarray}
\dot{t} & = & \dot{t}^{}_{\rm K} + L_z\,\delta g^{\rm CS}_{\phi}(r)\,,~~~~~~~~~~~\label{geoCS_1} 
\dot{\phi}  = \dot{\phi}^{}_{\rm K} - E\,\delta g^{\rm CS}_{\phi}(r)\,,\\ \nonumber		
\dot{r}^2  &=&  \dot{r}^2_{\rm K} + 2EL_zf\,\delta g^{\rm CS}_{\phi}(r)\nonumber~~~~~
\dot{\theta}^2  =  \dot{\theta}^2_{\rm K}\,, \label{dottheta}
\end{eqnarray}
where the dots denote differentiation with respect to proper time and $E$ and $L^{}_{z}$ are respectively the energy and the angular momentum per unit SCO mass. The quantities ($\dot{t}^{}_{\rm K}$, $\dot{r}^{}_{\rm K}$, $\dot{\theta}^{}_{\rm K}$, 
$\dot{\phi}^{}_{\rm K}$) represent the geodesic equations in GR, that is in Kerr, and the CS deviations are governed by the term $\delta g^{\rm CS}_{\phi}\propto \xi\cdot a$ ref.~\citep{Sopuerta:2009iy}. 

Since the orbit of the SCO is not a geodesic trajectory, but an inspiral that evolves adiabatically, the orbital parameters associated with the geodesic should be evolved to model the trajectory. To that end, we have developed an scheme to evolve the EMRI orbit in the framework of DCSMG ref.~\citep{Canizares:2012is} by incorporating radiation reaction (RR) effects, which not only drive the inspiral, but  also break the degeneracy between an orbit in GR and another in DCSMG. In other words, the RR pushes the SCO across the MBH geometry, which differs in both theories,  and hence, the GW generated in GR is dephased with respect to the one generated in DCSMG \citep{Sopuerta:2009iy}. Once the orbital evolution of the SCO is obtained, the waveform associated with its trajectory is computed through a multipolar expansion up to the mass quadrupole ref.~\citep{Thorne1980R}.
\begin{figure}[!h]
\begin{center}
\includegraphics[scale = 0.3, angle = 0]{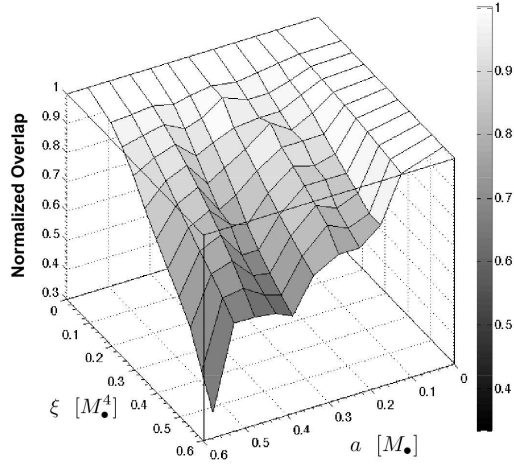}
\caption{Symmetric normalized overlap [Eq.~(\ref{o-gr-cs})], which shows the projection of a GW in DCSMG with the corresponding one in GR. \label{Fig1}}
\end{center}
\end{figure}


We describe EMRI systems with a $15$-dimensional parameter space including the CS parameter $\xi$: $\mb{\lambda} = \{M^{}_{\bullet}, a, \mu, e^{}_0, p^{}_0, \theta^{}_{\INC,0}, \zeta\equiv\xi\cdot a, \theta^{}_{\rm S}, \phi^{}_{\rm S}, \theta^{}_{\rm K}, \phi^{}_{\rm K} ,D^{}_{\rm L}, \psi^{}_0, \chi^{}_0, \phi^{}_0\}$ (see Table~\ref{table_parameters} for definitions), where the subscript $0$ refers to values taken at the initial time of the inspiral and $\psi$ and $\chi$ are angle variables (AV) introduced to describe the radial and polar motion respectively. However, the present work is restricted to a $5$-dimensional subset of the parameters that show the strongest correlations with $\xi$. The reduced parameter space is then given by $\mb{\lambda} = \{M^{}_{\bullet}, a, e^{}_0, \zeta, D^{}_{\rm L}\}$. Here, we assume $a /M_{\bullet}=e^{}_0=0.25$, and $\zeta/M^5_{\bullet} = 5\cdot10^{-2}$ and the rest of the parameters have fixed values given by: $ \mu=m^{}_{\star}/M_{\bullet}=10\,M^{}_{\odot}\,/M_{\bullet}$, $p^{}_{0} = 11\,$, $\theta^{}_{\INC,0}=0.569\,$, $\theta^{}_{\rm S} = 1.57\,$, $\phi^{}_{\rm S} = 1.57\,$, $\theta^{}_{\rm K}=0.329\,$, $D^{}_{\rm L} = 1\,$Gpc, and $\phi^{}_{\rm K}=\psi^{}_{0} = \chi^{}_{0}=\phi^{}_{0} = 0.78\,$.

\begin{table}
\centering
\small
\begin{tabular}{cc||cc}
\hline
\hline
Parameter                                    & ~Description					&Parameter                                    & ~Description \\
\hline
\hline
$M^{}_{\bullet}$                             & ~MBH mass [$M_{\odot}$]. 			&$\theta^{}_{\rm S}$                          & ~EMRI polar angle.\\

$a = |\mb{S}^{}_{\bullet}|/M^{}_{\bullet}$~  & ~MBH Spin [$M^{}_{\bullet}$].   &$\theta^{}_{\rm S}$                          & ~EMRI polar angle.\\
$\mu = m^{}_{\star}/M^{}_{\bullet} $ 	        & ~EMRI mass ratio. 			&$\theta^{}_{\rm K}$                          & ~MBH spin polar angle.\\

$e^{}_0$                                     & ~Eccentricity.&$\phi^{}_{\rm K}$                            & ~MBH spin azimuthal angle.\\
$p^{}_0$                                     & ~Semilatus rectum.&$D^{}_{\rm L}$                       & ~ Distance to the EMRI [Gpc].\\
$\theta^{}_{\INC,0}$                         & ~Orbit inclination .& $\psi^{}_0$                        & ~AV. for the radial motion.\\

									& &$\chi^{}_0$                                  & ~AV. for the polar motion.\\
$\zeta$                                      & ~$\xi\cdot a$~~[$M_{\bullet}^{5}$].& $\phi^{}_0$      & ~BL azimuthal angle.\\
\end{tabular} 
\caption{Summary of the parameters that characterize an EMRI system in DCSMG. Units are indicated in square brackets for parameters with dimension and $D_L$ is measured from the SSB to the EMRI. \label{table_parameters}} 
\end{table} 

In our analysis, we consider two different EMRI system, labelled $A$ and $B$, characterised by the mass of their MBH. In particular, for system $A$ we consider $M^{}_{\bullet}=5\cdot10^{5}M^{}_{\odot}$ and $M^{}_{\bullet}=10^{6}M^{}_{\odot}$ for system $B$. The parameter estimation errors expected for LISA for these systems are given in Table~\ref{table_error_estimates}. 

The signal-to-noise (SNR) for eLISA is around two times smaller than the one of LISA, then, to compare  the error estimates of both detectors we can normalize to a fixed SNR. The results obtained for system $A$ show that the parameter estimation accuracy does not change appreciably when the noise curve of LISA is changed for the one of 
eLISA and so, all previous results can be considered to apply to eLISA as well, with the corresponding SNR corrections ref.~\citep{Canizares:2012is}.

\begin{table}[ht]
\centering
\small
\begin{tabular}{c c} 
\hline
\hline
System A                &		System B\\
\hline
\hline
$\Delta\log M^{}_{\bullet} \sim 5\cdot10^{-3}$~~~&$\Delta\log M^{}_{\bullet} \sim 6\cdot10^{-4}$~~~~\\
$\qquad \Delta a\sim 5\cdot10^{-6}\,M^{}_{\bullet}$~~~~~~~~~~~~~~~~&$\qquad \Delta a\sim 3\cdot10^{-6}\,M^{}_{\bullet} $~~~~~~~~~~~~~~~~\\
$\Delta e^{}_{0}\sim 3\cdot10^{-7}$~~~~~~~~~~~~~&$\Delta e^{}_{0}\sim 10^{-7}$~~~~~~~~~~~~~~~~~ \\ 
$\Delta\log\,\zeta\sim 4\cdot10^{-2}$~~~~~~~~&$\qquad  \Delta\log\,\zeta\sim2\cdot10^{-2}$~~~~~~~~~~~~~~~~\\
$\Delta\log( D_L/\mu)\sim 2\cdot10^{-2}$ &$\Delta\log( D_L/\mu)\sim 2\cdot10^{-2}$\\
\hline
\hline
\end{tabular} 
\caption{Parameter estimation errors for LISA observations of the systems. \label{table_error_estimates}}
\end{table} 

In order to study how different a waveform in DCSMG, $\textbf{h}^{}_{\rm CS}$, is from the corresponding one in GR, $\textbf{h}^{}_{\rm GR}$, we can compute their (normalized) projection, i.e. their symmetric normalised overlap function:
\begin{eqnarray}
{\cal O}\left[\textbf{h}^{}_{\rm GR},\textbf{h}^{}_{\rm CS}\right]\equiv\frac{\left(\textbf{h}^{}_{\rm GR}|\textbf{h}^{}_{\rm CS} \right)}{\sqrt{\left(\textbf{h}^{}_{\rm GR}|\textbf{h}^{}_{\rm GR} \right)\left(\textbf{h}^{}_{\rm CS}|\textbf{h}^{}_{\rm CS} \right)}}\,, \label{o-gr-cs}
\end{eqnarray}
where both waveforms correspond to EMRI systems defined with the same parameters.

In Figure 1, we show how the projection of $\textbf{h}^{}_{\rm CS}$ onto $\textbf{h}^{}_{\rm GR}$ changes as the values of the MBH spin $a/M_\bullet$ and the CS parameter $\xi/M_{\bullet}^4$ are modified. These results are for system A, assuming a $0.5$yr EMRI observation. We see that, for higher values of $a/M^{}_{\bullet}$ and  $\xi/M^{4}_{\bullet}$ the overlap Eq.~(\ref{o-gr-cs}) decreases, since the difference in the evolution of the SCO in GR and CS increases and, consequently, the deviations of $\textbf{h}^{}_{\rm CS}$ from $\textbf{h}^{}_{\rm GR}$ are enhanced ref.~\citep{Canizares:2012is}. 

In our work ref.~\citep{Canizares:2012is}, we have performed numerical simulations to explore how the results change under different assumptions, namely the effects of (i) including RR, (ii) changing the value of the spin and (ii) changing the value of the CS parameter. The parameter estimation analysis is performed using the Fisher information matrix (FM) formalism ref.~\citep{Fisher:1935}, and in order to assess the validity of the FM results, we evaluate the maximum-mismatch criterion (MMC) ref.~\citep{Vallis2008}. Regarding (i), we evolved system $A$ with and without RR and for different evolution times: $T^{}_{\rm evol} = 0.1\,$, $0.3\,$, $0.5\,$, $1$ yr,  which ensures that enough high SNR  is accumulated to perform parameter estimation. We conclude that the inclusion of the RR improves the SNR of the signals and also the parameter estimates, in particular those of the spin, $a$, and of the CS parameter, $\zeta$. This is partially due to the increase of the overall SNR due to RR, but even after rescaling to a fixed reference SNR, we see an improvement in the parameter measurement accuracies when RR is included. As one could expect due to the adiabatic nature of the RR, the improvement with the inclusion of RR is more significant for longer evolution times. On the other hand, to  explore (ii), that is the effect of the MBH spin $a$ on the error estimates, we consider that systems $A$ and $B$ have the following values for the MBH spin: $a/M^{}_{\bullet} = 0.1$, $a/M^{}_{\bullet} = 0.25$, and $a/M^{}_{\bullet} = 0.5$. We found that the smaller the spin, $a/M^{}_{\bullet}$, the better the parameter estimate for the CS parameter, $\zeta$. In particular $\Delta \zeta\sim 2.8\cdot10^{-2}$ for system $A$ and 
$\Delta \zeta\sim 1.4\cdot10^{-2}$ for system $B$. The reason for this is that the CS modifications affect a single MBH metric component,
${\met}^{}_{t\phi}$ [Eq.~(\ref{DCSMG_metric})], which contains the CS parameter $\xi/M_{\bullet}^4$, multiplied by the spin parameter $a$. The unperturbed Kerr metric component is proportional to $a$, and so the relative change in this metric coefficient due to the addition of the DCSMG correction is proportional to $\xi$. Since we keep $\zeta= a\,\xi$ fixed as we vary $a$, the value of $\xi$ increases as $a$ decreases and so the CS correction to the MBH metric is larger relative to the leading order Kerr metric term. Finally, we address (iii) by varying the value of the CS parameter $\xi/M_{\bullet}^4$ of system $A$. In particular we consider: $\xi=0.05M_{\bullet}^4$, $\xi=0.1M_{\bullet}^4$ and $\xi=0.2M_{\bullet}^4$, finding that the value of the CS parameter $\xi/M_{\bullet}^4$ only affects significantly the error estimate of the CS parameter itself, whereas modifying $a/M^{}_{\bullet}$ has a major effect on the error estimates of all the parameters employed in our study, and in particular on $\zeta$.

With the framework that we have developed, we can go an step further and assess whether LISA could bound the value of the CS parameter  by estimating the error on the measurements of $\xi$, $\Delta\xi$, i.e. by setting a bound like $\xi < \Delta\xi$, assuming that GR is the true theory of gravity. As different EMRI systems will provide different constraints, and $\xi$ is a universal quantity, we just need to look for the EMRI system that provides the best constraint. In this way, we find that (e)LISA could place a bound $\xi^{1/4} <  1.4\cdot10^4\,$km, which is almost four orders of magnitude better than the bound obtained  in ref.~\citep{AliHaimoud:2011fw} using Solar System data (see ref.~\citep{Yagi:2012he} for a recent work where the CS parameter is estimated using BH binaries).

In summary, we have analysed how well a space-based GW detector like (e)LISA can discriminate between an EMRI system in GR and one occurring in a modified gravity 
theory like DCSMG. To that end, we have constructed a waveform template model in DCSMG and used the Fisher matrix formalism, together with the MMC, to
estimate errors in parameter measurements, finding that a GW detector like (e)LISA should be able to discriminate between the two theories with a fractional error below $5\%$. Furthermore, assuming that GR is the true theory of gravity, we have been able to constrain the CS parameter $\xi$ with bounds four orders of magnitude better than using Solar System data.


\bibliography{References_Canizares}
\bibliographystyle{asp2010}

\end{document}